\documentstyle [11pt] {article}
\makeatletter

\newcommand{\singlespacing}{\let\CS=\@currsize\renewcommand{\baselinestretch}{1}\tiny\CS}
\newcommand{\oneandahalfspacing}{\let\CS=\@currsize\renewcommand{\baselinestretch}{1.25}\tiny\CS}
\newcommand{\doublespacing}{\let\CS=\@currsize\renewcommand{\baselinestretch}{1.35}\tiny\CS}

\def\@citex[#1]#2{\if@filesw\immediate\write\@auxout{\string\citation{#2}}\fi
  \def\@citea{}\@cite{\@for\@citeb:=#2\do
    {\@citea\def\@citea{,\linebreak[0]\hskip0pt plus .2em}%
      \@ifundefined{b@\@citeb}%
      {{\bf ?}\@warning{Citation `\@citeb' on page \thepage\space undefined}}%
      \hbox{\csname b@\@citeb\endcsname}}}{#1}}

\newtheorem{rule-def}[theorem]{Rule}

\begin{document}

%\setcounter{chapter}{1}
% defining short form------
\newcommand{\la}{\lambda}
\newcommand{\si}{\sigma}
\newcommand{\ol}{1-\lambda}
\newcommand{\be}{\begin{equation}}
\newcommand{\ee}{\end{equation}}
\newcommand{\bea}{\begin{eqnarray}}
\newcommand{\eea}{\end{eqnarray}}
\newcommand{\nn}{\nonumber}
\newcommand{\lb}{\label}

\begin{center}
{\large\bf A Note On \\Astronomer R. G. Chandra and British
Astronomical Association}
\end{center}

\begin{center}

Sudhindra Nath Biswas$^{1}$, Utpal
Mukhopadhyay$^{2}$ \& Saibal Ray$^{3}$

$^{1}${\it Mahatma Aswini Kumar Dutta Road, Nabapally, Barasat, North 24
Parganas, Kolkata 700126, West Bengal, India}

$^{2}${\it Satyabharati Vidyapith, Nabapally, Barasat, North 24
Parganas, Kolkata 700126, West Bengal, India\\ umsbv@yahoo.in}

$^{3}${ \it Department of Physics, Government College of
Engineering \& Ceramic Technology, Kolkata 700010, West Bengal,
India \\ saibal@iucaa.ernet.in}\\

\end{center}

{\bf Abstract:} {\it In the present Note we have presented some
documents to reveal the longstanding relationship of Indian
amateur astronomer R. G. Chandra with British Astronomical
Association.}\\

\section{Introduction}
~~~~Radha Gobinda Chandra (1878 - 1975), an Indian amateur
astronomer, had a British connection not only in the sense that he
was then under the British Rule but also through his mentor
Kalinath Mukherjee, a law practitioner in the District Court of
Jessore, Bengal. While Mukherjee was a college student, he had the
opportunity to come in close contact with Sir M. J. Herschel, M.
A. and Bar-at-Law who was the then District and Session Judge of
Nadia district of undivided Bengal. He was the son of Sir John
Herschel (1792 - 1871) and the grandson of Sir William Herschel
(1738 - 1822) who were the two pioneer astronomers$^{1}$.

However, in the present note we would like to investigate
Chandra's direct British connection through his scientific works,
especially with the British Astronomical Association (BAA), who
elected Chandra as their honorary member for his important
contributions in observational astronomy.\\

\section{Observation Of Comets}

\subsection{7P/Pons-Winnecke}
~~~~The first British connection of Chandra can be traced back
through the reporting ``Search for meteors from the Pons-Winnecke
Radiant", in the journal {\it Nature}$^2$ which goes as: {\it R.
G. Chandra of Jessore, India, also reports a fruitless search for
meteors in the night of June 25. He states that Prof. Ray of
Bolpur saw two meteors radiating from the neighbourhood of
$\theta$ Bootes.}\footnote{ Here, Prof. Ray means Jagadananda Roy
(1869 - 1933) of Santiniketan who was a renowned science teacher
of the school at Santiniketan under Viswabharati University,
Bolpur founded by the Nobel Laureate poet Rabindranath Tagore
(1861 - 1941).}

Chandra was successful in observing the comet $7P/Pons-Winnecke$.
It is known that on 20 June 1927, at about $15.5^h$ U. T. ($21^h$
IST), Chandra was busy with his usual scheduled programme for
observations of variable stars. He suddenly noticed a nebula-like
object just North-West of the bright star Vega ($\alpha$ Lyrae).
At that time the celestial body was visible on the line joining
star $\gamma$ Draconis and $\alpha$ Lyrae and was nearer (RA:
$18^h~22^m~30^s$, Dec: $+ 40^0~30^{\prime}$) to the latter one.
After consulting the handbook of BAA, he came to know that the
object under his observation was the comet $7P/Pons-Winnecke$.
During the period of his observation, Chandra observed the comet
to pass through the constellations Lyra, Cygnus, Vulpecula,
Delphinus, Pegasus, Acquarius, Sculptor and Phoenix at a very fast
speed of $40,000$ km/hr.

\subsection{2P/Encke}
~~~~Following the instruction of A. C. D. Cromlin, the Director of
BAA, Chandra$^3$ also observed the comet $2P/Encke$ 1927 having
shortest period of 3.30 years among the periodic comets. He
searched out this comet in 1928 at 7 PM from Jessore with his
3-inch telescope in the Pegasus. This comet which appeared to him
as a small nebulosity near the Andromeda galaxy (M31) remained
visible from 19 October 1927 to 3 April 1928 during the
apparition.

\subsection{C/1942~X1~Whipple-Fedtke-Tevzadze}
~~~~On 24 February 1943, at about $16.5$~U. T. ($22.00$ IST)
Chandra$^3$ noticed a nebula-like object near the star $\gamma$
Ursae Majoris. Later he could recognize the object as a comet
though could not identify it. He made series of observations on
the comet until 10 May 1943 with the help of two refracting
telescopes, one of his own 3-inch and the other 6.25-inch lent to
him by the American Association of Variable Star Observers
(AAVSO). Chandra recorded (i)~the~apparent~path and cometary
phenomena of the comet, and (ii) the abrupt variations in its
brightness. Later on, he came to know from the Journal of BAA that
the long period comet $C/1942~X1~Whipple-Fedtke-Tevzadze$ was
really a variable one and the phenomena of variations in magnitude
was due to influence of solar magnetic disturbances during a
sunspot maximum.\\

\section{Variable Star Observer}
~~~~As an honorary member of BAA and AAVSO, Chandra's
responsibility was to contribute his collected observational data
on the period and magnitude of the variable stars. Since 1919, his
observational data were regularly published in the Monthly Report
of AAVSO$^4$, Memoirs of the BAA$^5$ and elsewhere. In this way he
made a good relationship with different scientific personalities,
like Felix de Roy, Director, Variable Star Section, BAA who in a
letter (dated 27 December 1923) expressed his desire to meet
Chandra, by writing: {\it ``I seize this opportunity in saying
that your excellent observations and remarks are always much
valued by this section. ... I should be personally pleased to meet
you if ever you were to cross to Europe."}$^6$

Probably in this same spirit Leon Campbell, Recording Secretary of
AAVSO has remarked in his article `The Role of the Amateurs in
Variable Star Astronomy' that {\it ``In foreign countries we have
Radha G. Chandra, official of Bagchar, India. Mr. Chandra, now in
his sixtieth year, who has been aiding in the variable star work
since 1919, has accumulated probably more observations on variable
stars than any other AAVSO foreign observer, well over
50,000."}$^6$

So it is really understandable that how difficult is it to collect
and represent the huge number of observations made by Chandra
throughout his life. Actually, in the year ending in October 1926
he made no fewer than 1685 observations, of which he made 226
observations in the month of March only$^7$. Chandra reported 34
individual variable stars to the BAA$^8$ on which he made
observations during the period 1920-24 (Table-1).

\begin{table}
\caption{Observations made by Chandra on 34 variable stars (1920 -
1924)}
\centering
\bigskip
{\small
\begin{tabular}{@{}llrrrrlrlr@{}}
\hline \\[-9pt] VARIABLE OBSERVED & NO. & VARIABLE OBSERVED &
NO.\\ \hline \\ [-6pt]

R Andromaedae &  21& V Cygni   &14\\

W Andromedae &55& R Draconis &16\\

R Aquilae &7 & T Draconis &3\\

R Arietis &42& R Geminorum &33\\

R Aurigae   &25& S Herculis &3\\

X Aurigae  &50&T Herculis &34\\

R Bootis  &65&U Herculis &45\\

S Bootis &53&R Hydrae &101\\

V Bootis &57&R Leonis &33\\

R Camelopardalis &21&U Orionis &48\\

X Camelopardalis &29&R Pegasi &7\\

T Cassiopeiae &39&R Serpentis &10\\

T Cephei &17&V Tauri &18\\

s (Mira) Ceti &136&R Ursae Majoris &30\\

S Coronae &62&S Ursae Majoris &69\\

X Cygni &86&T Ursae Majoris &48\\

R Cygni &19&S Virginis &52
\\

\hline
\end{tabular}
}
\end{table}

\section{Observation Of Lunar Eclipse}
~~~~Chandra observed the occultation and lunar eclipse of 20
February 1924 which he reported to BAA$^9$ as follows: {\it
``Arrangements were made with two friends to observe the Lunar
eclipse and occultations of stars, one to watch the minute hand,
the other to watch the second hand, and both counting the minutes
and seconds independently and record the time when I shouted
`one', `two' and `three' from the telescope. This was carefully
done so that we get a very accurate time. Time was taken from the
Jessore Telegraph Office at 4 p.m. at which hour each day the time
is signaled from the Government Telegraph Office at Calcutta. The
sky was very fine and seeing very good: the observations were made
with naked eye, with binoculars and with a 3-inch refractor using
powers of 32 and 80.''}

The observational report of the Lunar eclipse of 26 September 1931
was published in the Vol. $42$ of {\it The Journal of the British
Astronomical Association}$^{10}$ and observation of Annular Solar
Eclipse of 21 August 1933 was also reported in the same
journal$^{11}$ in the Vol. $44$. BAA published the report of the
Annular Solar Eclipse in the following form :

{\it Observations of Annular Eclipse of the Sun at Jessore
(India), 1933, August 21\\
 Latitude $23$ deg. $10$ min. $5$ sec., Longitude $89$ deg. $15$
 min. $15$ sec. \\
First Contact - $3^h$$40^m$$15^s$ GMT.\\ Second Contact (Formation
of Annulus) - $5^h$$19^m$$59.2^s$ GMT.\\ Last contact not
observable owing to clouds.\\ Instrument : $3$-inch refractor\\ An
attempt was made to see the corona by putting the Sun out of the
field of view, but without success.}

A short communication by Chandra on `Rahu' was published in the
above mentioned journal$^{12}$ in a different issue. The
communication is reproduced below :

{\it Rahu - There have been several communications to the
Association [JBAA 43, 317 (1933); JBAA 45, 322 (1935)] regarding
`Rahu', but so far there has been no correct explanation. Hindoo
astronomy may be divided into two sections, mythological and
mathematical. In the Indian Epics there are many narratives and
fables relating to celestial bodies, and these may be taken to
constitute the first of these sections. As well as Rahu there is
another term Ketu, and there are with these, planets called
`Nabagraham', viz., Rabi, the Sun; Shome, the Moon; Kuja, the
Mars; Buddha, Mercury; Guru, Jupiter; Shitau, Venus; Manda,
Saturn, Rahu being the ascending node and Ketu the descending
node. These are regarded as powerful deities having influence on
the affairs of humnity, and they are also used in Indian
astrology.

In the Sreemat Bhagabat Puran, Part 8, Chapter 9, Shloks 21-23, it
is narrated that Rahu was a Danab (demon), the son of Shinghika,
the wife of Biprachitti. The Danabs were the antiparty of the
Devs, having no right to drink the `Amrita', a divine liquor which
made the Devs immortal. Rahu made an attempt in disguise to drink
the Amrita along with the Devs, but this was pointed out by Surja
(the Sun) and Chandra (the Moon). On this, Hari, the Prince of the
Devs, cut off the head of Danab Rahu, but as a small quantity of
the Amrita entered the throat his head became immortal and was
placed in the sky as a `Graha'. The Sun and Moon having betrayed
his attempt to drink the Amrita, Rahu developed a hatred for them,
which has resulted in his endeavouring to devour them whenever he
gets an opportunity; this he cannot do because he has only a head
and no body.

Thus is the story, from the epic, of Rahu and the cause of
eclipses; but Indian astronomers have known the real causes from
time immemorial.

As Rahu's body was not immortal it was thrown away, but at a later
date Indian astronomers and astrologers placed the body, as Ketu,
$180^0$ from Rahu in the Zodiac.

In Hindoo mathematical astronomy Rahu and Ketu, the ascending and
descending nodes of the planet's orbits, resemble the Greek
Dragon's Head and tail, and are the two points where eclipses
ocur, being referred to as Patha in the Surja Siddhanta. - R. G.
Chandra.}

\section{Conclusion}
Besides Felix de Roy we would like to mention here some other
astronomers of BAA with whom a cordial relationship were built by
Chandra. Here is a part of a letter (dated 28 August 1928),
written by A. N. Brown of BAA to be quoted:

{\it ``I acknowledge with thanks the report of 15 complete sheets
of your observation of variables made this year ... I have so far
only just glanced over your sheets, but this glance is sufficient
to show that you have again done valuable work, particularly,
perhaps, in the regularity of your observations of some of the
Irregular $U$ Germinorum etc. in spite of the unfavourable weather
with which you say you have had ..."}

From another letter by Y. M. Holborn, the then secretary of BAA,
we come to know how indispensable were, for the professionals, the
data collected by Chandra from the observation of variable stars.

Due to his very old age when Chandra tendered his resignation from
the membership of BAA, then Holborn reacted as follows (letter
dated 30 January 1941):

{\it ``I am passing on your letter of resignation to Mr. Brown who
deals with these things. But I must say, I think it is a great
pity to resign at this time when the Association is in the utmost
need of support.

Your longstanding work for the variable star section too will be
greatly missed just at the time when Lindley and others like
myself with full time war duties have had to give up observing.

I beg to you as an old member to reconsider this decision of
yours."}

However, no concrete official document is available to us to
ascertain whether Chandra ultimately withdrew his resignation or
not. We guess he did not as very sincere Chandra thought that
without contributing any astronomical data to BAA he should not be
an ornamental member. Not only that, when old age forced Chandra
to withdraw himself from active astronomoical activities, he
donated all his books and periodicals (sent to him by BAA, AAVSO
and other organizations) as well as his legendary $3$-inch
telescope to Barasat Satyabharati Vidyapith, a Higher Secondary
School close to his residence, for the use of future generations.
Finally, it may be mentioned that some volumes of those Journals
are still there in the school library of that institution which
is, incidentally, the past and present working places of the first
and second author respectively.\\

\vspace{2.5cm}

\begin{center}
 REFERENCES
\end{center}

\begin{enumerate}

\item{} S. N. Biswas, U. Mukhopadhyay and S. Ray, arXiv:physics.hist-ph/0195697
(2011).

\item{} {\it Nature}, Vol. 107, No. 2700, p. 694 (1921).

\item{} Radhagobinda Chandra, {\it Dhumketu} (Puthipatra, Kolkata, 1985).

\item{} {\it Monthly Reports and Annual Reports of the American
Association of Variable Star Observers}, p. 133 (1926).

\item{} {\it Memoirs of the British Astronomical Association}, vol. XXVIII, Table C. (1929).

\item{} Ranatosh Chakrabarty, {\it Jyotirbijnani Radhagobinda}
(Puthipatra, Kolkata, 1999).

\item{} Rajesh Kochhar and Jayant Narlikar, {\it Astronomy in India:
Past, Present and Future} (IUCAA, Pune and IIA, Bangalore, 1993).

\item{} Otto Struve and Velta Zebres, {\it Astronomy in the 20th
Century} (Macmillan Co., New York, p. 354, 1962).

\item{} {\it Journal of the British Astronomical Association} 34, 241 (1924).

\item{} Ibid, 42, 178 (1931).

\item{} Ibid, 44, 157 (1934).

\item{} Ibid, 45, 407 (1935).

\end{enumerate}

\end{document}